\documentclass[fleqn,usenatbib]{mnras}

\usepackage{newtxtext,newtxmath}

\usepackage[T1]{fontenc}

\DeclareRobustCommand{\VAN}[3]{#2}
\let\VANthebibliography\thebibliography
\def\thebibliography{\DeclareRobustCommand{\VAN}[3]{##3}\VANthebibliography}

\usepackage[dvips]{graphicx}	

\def\Msun{$M_{\rm \odot}$}
\def\ps{$\rm km \,s^{-1}\,kpc^{-1}$}

\title[Age distribution in BPX bulges formed without bar buckling]
{Age distribution of stars in boxy/peanut/X-shaped bulges formed without bar buckling}
\author[J. Baba et al.]{
Junichi \textsc{Baba}$^{1,2}$\thanks{E-mail: jun.baba@nao.ac.jp; babajn2000@gmail.com (JB); d.kawata@ucl.ac.uk (DK)}, 
Daisuke \textsc{Kawata}$^{3,1}$,
and
Ralph \textsc{Sch{\"o}nrich}$^3$
\\
$^1$ National Astronomical Observatory of Japan, Mitaka, Tokyo 181-8588, Japan.\\
$^2$ SOKENDAI (The Graduate University for Advanced Students), Shonan Village, Hayama, Kanagawa 240-0193, Japan.\\
$^3$ Mullard Space Science Laboratory, University College London, Holmbury St. Mary, Dorking, Surrey, RH5 6NT, UK.\\
}

\begin{document}

\date{Accepted xxx. Received xxx; in original form XXX}

\maketitle

\begin{abstract}
{Some barred galaxies}, including the Milky Way, host a boxy/peanut/X-shaped bulge (BPX-shaped bulge). Previous studies suggested that the BPX-shaped bulge can either be developed by bar buckling or by vertical inner Lindblad resonance (vILR) heating without buckling. In this paper, we study the observable consequence of an BPX-shaped bulge built up quickly after bar formation via vILR heating without buckling, using an $N$-body/hydrodynamics simulation of an isolated Milky Way-like galaxy.
We found that the BPX-shaped bulge is dominated by stars born prior to bar formation. This is because
the bar suppresses star formation, except for the nuclear stellar disc (NSD) region and its tips. The stars formed near the bar ends have higher Jacobi energy, and when these stars lose their angular momentum, 
{their non-circular energy}
increases to conserve Jacobi energy. This prevents them from reaching the vILR to be heated to the BPX region.
By contrast, the NSD forms after the bar formation. 
From this simulation and general considerations, we expect that the age distributions of the NSD and BPX-shaped bulge formed without bar buckling do not overlap each other.
Then, the transition age between these components betrays the formation time of the bar, and is testable in future observations of the Milky Way and extra-galactic barred galaxies.

\end{abstract}

\section{Introduction}
\label{sec:Introduction}

The near-infrared images from the COBE satellite presented the first clear evidence of the boxy/peanut-shaped bulge in the Milky Way \citep{Weiland+1994,Dwek+1995,Binney+1997}. Recent star counts have sharpened this picture by proving the presence of an X-shaped bar \citep[][]{McWilliamZoccali2010,Saito+2011,Ness+2012,Nataf+2015,WeggGerhard2013,Portail+2015b,NessLang2016}.
Using the red clump stars' magnitude distributions, \citet{Wegg+2015} fully established the full picture of a boxy/peanut/X-shaped bulge (BPX-shaped bulge) in the Milky Way.
The BPX-shaped bulge has a vertically extended structure (up to about 1 kpc) with the radial extension of around 1.5 kpc along the major-axis, which is an additional inner structure to a long and thinner Galactic bar with a length of about 5 kpc in radius and a scale height of about 180 pc \citep[see][for a review]{Bland-HawthornGerhard2016}.
The BPX-shaped bulge morphology is not unique to the Milky Way and such bulges are observed in external disc galaxies. The fraction of galaxies with BPX-shaped bulges is about half of the nearby edge-on disc galaxies \citep[][]{Lutticke+2000b,Lutticke+2004,Laurikainen+2014}. This fraction also strongly depends on mass \citep{ErwinDebattista2017,Li+2017} and declines towards higher redshift \citep{Kruk+2019}. 

{The origin of the BPX-shaped bulges} is closely related to bar formation \citep[][]{SellwoodWilkinson1993,SellwoodGerhard2020}. One of physical mechanisms about the formation of BPX-shaped bulge is buckling instability \citep{Toomre1966}, 
which is a common phenomenon in collisionless $N$-body simulations of disc galaxies \citep{Raha+1991,FriedliPfenniger1990,PfennigerFriedli1991,MerrittSellwood1994,Martinez-ValpuestaShlosman2004,Martinez-Valpuesta+2006,Debattista+2006,Debattista+2017,Debattista+2018,Debattista+2020,Fragkoudi+2017,Saha+2013,SmirnovSotnikova2018,SmirnovSotnikova2019,Lokas2019,Khoperskov+2019,Collier2020}.
The buckling instability involves spontaneous {breaking} of the symmetry with respect to the disc equatorial plane, developed by the vertical inner Lindblad resonance \citep[vILR, e.g.][]{PfennigerFriedli1991},
that thicken and weaken the bar on a few dynamical timescales. Numerical simulation studies showed that the buckling event occurs of order one to a few Gyr after the bar forms.\footnote{Note that subsequent buckling events can happen at a later time {under certain conditions} \citep{Martinez-Valpuesta+2006,Saha+2013,SmirnovSotnikova2019}.}
Some previous studies \citep{Shen+2010,GerhardMartinez-Valpuesta2012} showed that the buckled bar naturally reproduced the observed BPX-shaped properties of the Milky Way in many aspects \citep[][for a review]{LiShen2015,ShenLi2016}. 
However, it should be noted that the presence of gas suppresses buckling of bars, as shown in previous $N$-body/hydrodynamics simulations \citep{Berentzen+1998,Debattista+2006,Berentzen+2007,WozniakMichel-Dansac2009,Villa-Vargas+2010}. 

Other mechanisms invoked to explain the formation of BPX-shaped bulges are vertical resonant heating \citep[][]{CombesSanders1981,Combes+1990,FriedliPfenniger1990,PfennigerFriedli1991,Quillen+2014} or resonant trapping into a vILR secularly during bar growth \citep{Quillen2002,SellwoodGerhard2020}.
Recently, using $N$-body simulations, \citet{SellwoodGerhard2020} demonstrated that these two mechanisms can develop the BPX-shaped bulge without bar buckling soon after the formation of the bar.

The vertical resonant heating without bar buckling was seen in many previous simulations, in some of which the heating mechanism was not explicitly mentioned. These studies showed that a planar orbit family of the bar \citep[i.e. $x_1$ orbits;][]{ContopoulosPapayannopoulos1980} bifurcates into a 3D orbit family, $x_1v_1$ orbits \citep[so-called banana orbits;][]{PfennigerFriedli1991,Patsis+2002,Skokos+2002a,Williams+2016}, or higher-order resonant orbits such as brezel-like orbit families \citep{Portail+2015b,Valluri+2016,Abbott+2017,PatsisHarsoula2018,Parul+2020}.
In addition to these studies of stellar orbits in barred potentials, \citet{WozniakMichel-Dansac2009} performed $N$-body/hydrodynamics simulations including star formation, and showed that, without bar buckling, stars born in the gaseous disc rapidly populate vertically resonant orbits triggered by the combined effects of the horizontal ILR (hILR) and vILR.

In this mechanism of the BPX-shaped bulge formation without buckling, the BPX-shaped bulge starts forming soon after the bar forms, confined in radius by the extent of the vILR. 
Interestingly, previous numerical simulations also suggested that during the bar growing phase, intense star formation in the central sub-kpc region forms a nuclear stellar disc \citep[NSD; e.g.][]{FriedliBenz1993,FriedliBenz1995,MartineFriedli1997,HellerShlosman1994,Athanassoula2005,Wozniak2007,KimSaitoh+2011,Cole+2014,Debattista+2015,Debattista+2018,Seo+2019,BabaKawata2020a}, while the bar supresses star formation throughout the remainder of its extent \citep[][]{MartineFriedli1997,Spinoso+2017,Khoperskov+2018,Donohoe-Keyes+2019}. 
Consequently, after bar formation, no more stars should be born in the typical radial range of the vILR. While there is usually enhanced star formation in a ring around the bar, e.g. the Milky Way's 4 kpc ring, there is no known viable mechanism to transfer these stars inwards to the vILR. As a consequence, we can expect a marked age disparity between the BPX-shaped  bulge and the NSD: the BPX-shaped bulge is dominated by stars formed before the bar formation, while the NSD is dominated by the stars younger than the age of the bar \citep{BabaKawata2020a}.

Using an $N$-body/hydrodynamics simulation of an isolated Milky Way-like barred galaxy, where buckling is {\it not} suppressed artificially, we demonstrate that this is indeed the case, when the BPX-shaped bulge forms without buckling. 
In Section \ref{sec:ModelMethod}, we describe our galaxy model and simulation method. 
We analyse the morphological evolution of the simulated galaxy and quantify the time evolution of the simulated BPX-shaped bulge in Section \ref{sec:Morphology}.
In Section \ref{sec:Rbirth_tbirth}, we describe differences of the age distributions of the stars in the NSD, the in-plane bar structure (sometimes coined the long bar in the Milky Way), and the BPX-shaped bulge. 
In addition, Section~\ref{sec:Orbit} analyses orbital characteristics of stars formed after bar formation to explore what prevents these populations from being heated to the BPX-shaped bulge.
Finally, we summarise our results in Section \ref{sec:Conclusions}. 
Note that we consider only the case of the BPX-shaped bulge formation without bar buckling. We discuss the comparison with bar buckling driven BPX-shaped bulge briefly in Section \ref{sec:Conclusions}. 

\section{Models and Method}
\label{sec:ModelMethod}

\begin{figure*}
\begin{center}
\includegraphics[width=0.95\textwidth]{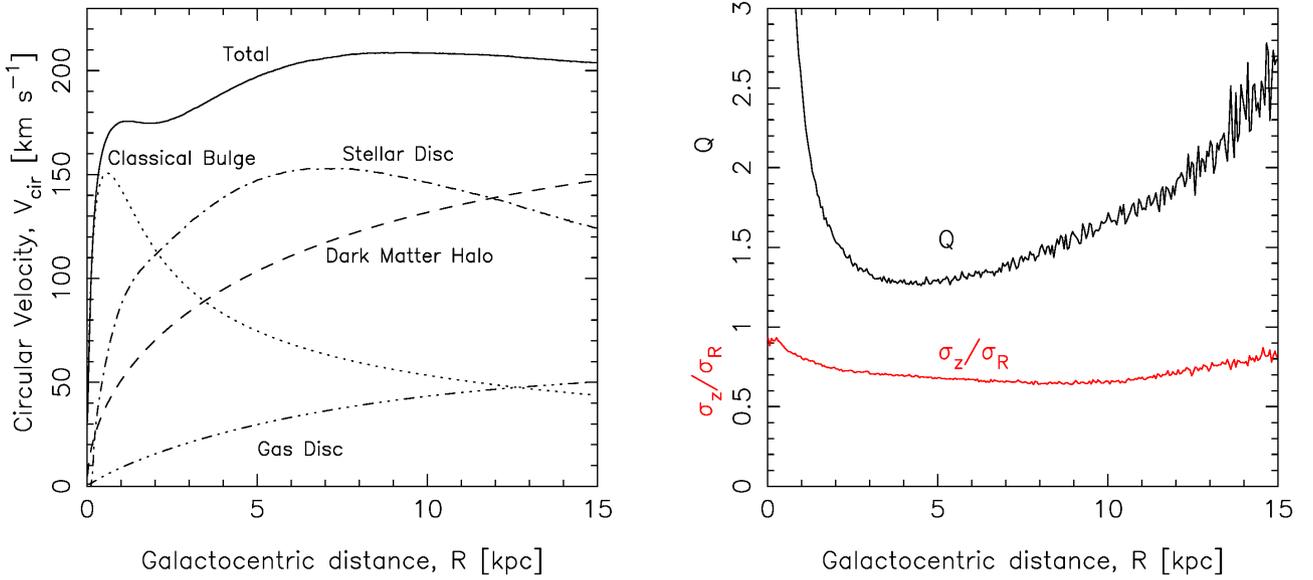}
\caption{
    (left) Initial circular velocity curves of each component of the galaxy model.
    (right) Initial Toomre's $Q$ value and {$\sigma_z/\sigma_R$} of the disc stars as a function of $R$.
}	
\label{fig:IC}
\end{center}
\end{figure*}

For the purpose of this study, we analyse an $N$-body/hydrodynamic simulation of an isolated galactic disc presented in \citet{BabaKawata2020a}. {Since we kept running the simulation used in \citet{BabaKawata2020a} up to $t=7$~Gyr, we present the results based on the simulation up to $t=7$~Gyr rather than $t=5$ Gyr available at the time of the publication of \citet{BabaKawata2020a}. }
In this section, we give a brief overview of the galaxy model and simulation techniques.

{We assume that the galaxy is initially composed of axisymmetric stellar/gas discs,  classical bulge and dark matter (DM) halo \citep[see also][for details]{Baba2015c}.}
The stellar disc follows a radially exponential and vertically isothermal profile:
\begin{eqnarray}
 \rho_{\rm \ast}(R,z) = \frac{M_{\rm d}}{4\pi R_{\rm d}^2 z_{\rm d}} 
 \exp\left(-\frac{R}{R_{\rm d}}\right){\rm sech}^2\left(\frac{z}{z_{\rm d}}\right), 
\end{eqnarray}
where $M_{\rm d}$, $R_{\rm d}$ and $z_{\rm d}$ are the total mass, scale-length and scale-height of the stellar disc, respectively. We assume that $M_{\rm d} = 4.3 \times 10^{10}~M_\odot$, $R_{\rm d} = 2.6$ kpc and $z_{\rm d} = 300$ pc \citep{Bland-HawthornGerhard2016}. Using Hernquist's method \citep[]{Hernquist1993}, the velocity structure of the stellar disc in cylindrical coordinates is determined by a Maxwellian approximation. The radial velocity dispersion is set to be Toomre's $Q=1.3$ at $R = 2.5R_{\rm d}$. The gaseous disc also follows an exponential profile with a total mass ($M_{\rm g}$) of $1.2 \times 10^{10}~M_\odot$, a scale-length of $10.4$ kpc and a scale-height of $100$ pc \citep[e.g.][]{BigielBlitz2012}. The initial temperature is set to $10^4$ K. The classical bulge follows the Hernquist profile with an isotropic velocity dispersion \citep{Hernquist1990}:
\begin{eqnarray}
 \rho_{\rm cb}(r) = \frac{M_{\rm b,0}}{2\pi}\frac{a_{\rm b}}{r(r+a_{\rm b})^3},
\end{eqnarray}
where $M_{\rm b,0}$ and $a_{\rm b}$ are the total mass and scale-length of the bulge, respectively. We assume that $M_{\rm b,0} = 2 \times 10^{10}~M_\odot$ and $a_{\rm b} = 0.79$ {kpc}. Following \citet{WidrowDubinski2005}, we generate the classical bulge using a distribution function with an energy cutoff with $q_{\rm b} = 0.21$ in equation~(11) of \citet{WidrowDubinski2005}. {The} resulting mass of the classical bulge ($M_{\rm b}$) is $6.7 \times 10^9~M_\odot$. 
{
As a result, the mass ratio of the classical bulge to the stellar disk is about 15\%. This value is somewhat larger than the current upper limit of the classical bulge mass fraction for the Milky Way 
\citep[$\lesssim$10\%;][]{Shen+2010,DiMatteo+2014,Debattista+2017}.
}

The initial numbers of stars and gas (SPH) particles are $5.7$ millions and $4.5$ millions, respectively, and particle masses for star and gas particles are about $9.1 \times 10^3~M_{\rm \odot}$ and $3 \times 10^3 ~M_{\rm \odot}$, respectively. 
{
In our simulations, the gas mass fraction $M_{\rm g}/(M_{\rm d}+M_{\rm b})$ is initially about 24\%. Because the scale-length of the gaseous disc is large, the gas surface density at $R = 8$ kpc is about $10~\rm M_\odot~pc^{-2}$. This is consistent with the observational value of the Milky Way \citep[e.g.][]{McKee+2015}. 
}

{
We model the DM halo with a rigid background potential. For the rigid DM halo, we adopt the Navarro-Frenk-White profile \citep{Navarro+1997}:
\begin{eqnarray}
 \rho_{\rm h}(r) =  \frac{M_{\rm h}}{4\pi f_{\rm c}(C_{\rm h})}\frac{1}{r(r+a_{\rm h})^2},
\end{eqnarray}
where $M_{\rm h}$, $a_{\rm h}$ and $C_{\rm h}$ are the total mass, scale radius and concentration parameter of the dark matter halo, respectively, and $f_{\rm c}(C_{\rm h}) = \ln(1+C_{\rm h}) - C_{\rm h}/(1+C_{\rm h})$. We assume that $M_{\rm h} = 1.26 \times 10^{12}~M_\odot$, $a_{\rm h} = 25$ kpc and $C_{\rm h} = 11.2$. 
Note that a rigid DM halo omits dynamical friction on the bar and suppress the slowdown of the bar \citep[e.g.][]{DebattistaSellwood2000,AthanassoulaMisiriotis2002}. Hence, our simulation does not explicitly include the slowdown of the bar. 
}

The $N$-body representation of the disc galaxy is initialised using the procedure described in \citet{Hernquist1993}. This method does not provide a strictly equilibrium model.
We resolve this by evolving the stellar orbits self-consistently for 6 Gyr, while fixing the positions of the gas particles and enforcing axisymmetry of the potential to prevent structure formation \citep{McMillanDehnen2007}.
We use this equilibrium state as the `initial' condition (i.e. $t = 0$ Gyr). Fig.~\ref{fig:IC} shows the initial circular velocity of each component (left panel) and Toomre's $Q$ value of the disc stars (right panel) as functions of the galactocentric distance, $R$.


Our simulation is carried out with an $N$-body/smoothed particle hydrodynamics (SPH) simulation code, {\tt ASURA-2} \citep{SaitohMakino2009,SaitohMakino2010}. Gravitational interactions of stars and SPH particles are calculated by the Tree with GRAPE method \citep{Makino1991}, using a software emulator of GRAPE, known as Phantom-GRAPE \citep{Tanikawa+2013}\footnote{\url{https://bitbucket.org/kohji/phantom-grape}.}. A gravitational softening length is set to $10$ pc in our simulation, and is sufficiently small to resolve the three-dimensional structure of a disc galaxy \citep{Baba+2013}.
The simulations also take into account radiative cooling for a wide temperature range of $20~{\rm K} < T < 10^8~{\rm K}$ \citep{Wada+2009}, heating due to far-ultraviolet interstellar radiation \citep{Baba+2017}, probabilistic star formation from the cold dense gas \citep[$T<100~\rm K$ and $n > 100~\rm cm^{-3}$;][]{Saitoh+2008}, as well as thermal feedback from type II supernovae \citep{SaitohMakino2009} and $\rm H_{II}$ regions \citep{Baba+2017}. 
{
To compensate for gas consumption due to star formation, the SPH particles are continuously added with a constant rate of $2~\rm M_\odot~yr^{-1}$, which models the gas accretion from the halo to the disc \citep{BabaKawata2020a}.
At $t = 7$ Gyr, the gas mass is about $1.5\times10^{10}~\rm M_\odot$ and the stellar mass has increased by about $1.0\times10^{10}~\rm M_\odot$ due to star formation. The final gas surface density at $R = 8$ kpc is about $10.7~\rm M_\odot~pc^{-2}$. Therefore, the gas mass is almost constant from the beginning of the simulation, while the total stellar mass has increased by about 20\%.}

As we aim to investigate {the} BPX-shaped bulge formation without buckling, we {choose the initial {conditions} so that the bar buckling is suppressed}. The left panel of Fig.~\ref{fig:IC} shows that the contribution from the stellar disc to the total circular velocity in the galaxy model is larger than that from the DM halo in the regions of $3 \lesssim R \lesssim 12$ kpc. Early bar formation is ensured, since the initial stellar distribution satisfies the criteria for the bar instability in a rigid DM halo \citep[e.g.][]{Efstathiou+1982}. However, the buckling instability happens if the stellar vertical-to-radial velocity dispersion ratio, {$\sigma_z/\sigma_R$}, is less than about 0.3 \citep[][]{Sellwood1996}\footnote{This is not a strictly correct condition for buckling. \citet{MerrittSellwood1994} discussed that the  buckling modes are maintained when $\Omega_z>2 (\Omega_\phi-\Omega_{\rm b})$ \citep{Debattista+2017}, where $\Omega_z$, $\Omega_\phi$ and $\Omega_{\rm b}$ are the vertical oscillation and mean angular frequencies of a star and the pattern speed of the bar, respectively. However, because $\Omega_{\rm b}$ is not known at the initial condition, we use this simpler empirical condition.}. 
As seen in the right panel of Fig.~\ref{fig:IC} our galaxy is set up to achieve an initial {$\sigma_z/\sigma_R>0.6$}. We additionally ensure that the initial bulge-to-disc mass ratio of $M_{\rm b}/M_{\rm d} = 0.16$ also exceeds the limit for bar buckling established by \cite{SmirnovSotnikova2019}. 
Also, as mentioned in Section~\ref{sec:Introduction} including the gas component and star formation further suppresses bar buckling.

\section{BPX-shaped bulge formation without bar buckling}
\label{sec:Morphology}

\begin{figure*}
\begin{center}
\includegraphics[width=0.95\textwidth]{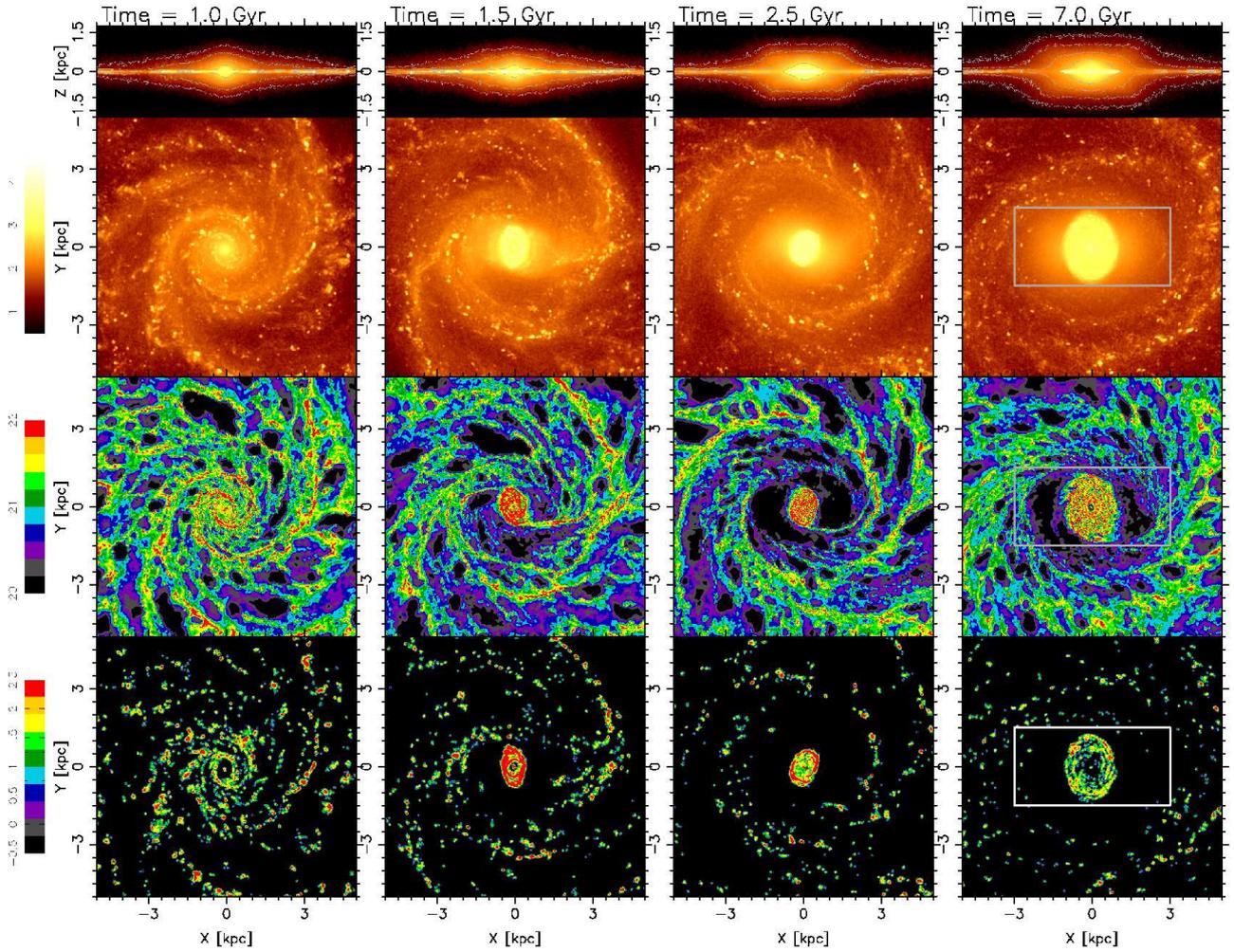}
\caption{
	Morphological evolution of the simulated barred spiral galaxy (the galaxy rotates in clock-wise {direction}). 
	Top row: Evolution of edge-on views. Orange colors indicate surface density of stars in logarithmic scale ($\rm M_\odot~pc^{-2}$). 
	{
	2nd row: Evolution of face-on views of the stellar surface density in logarithmic scale. 
	3rd row: Evolution of face-on views of the gas surface density in logarithmic scale ($\rm H~pc^{-2}$).
	}
	Bottom row: Evolution of star-formation-rate (SFR) distributions in logarithmic scale (\Msun{}$\rm~pc^{-2}~Gyr^{-1}$). Here, we computed the SFR using the stars younger than 10 Myr at each snapshot.
	After the bar formed ($t \gtrsim 1.5$ Gyr $\equiv T_{\rm bar}$), the major-axis of the bar is set to be the $x$-axis.
	In this study, we focus on the bar/bulge region enclosed by rectangle in the left panels. 
}	
\label{fig:SnapshotEvolution}
\end{center}
\end{figure*}

\begin{figure}
\begin{center}
\includegraphics[width=0.45\textwidth]{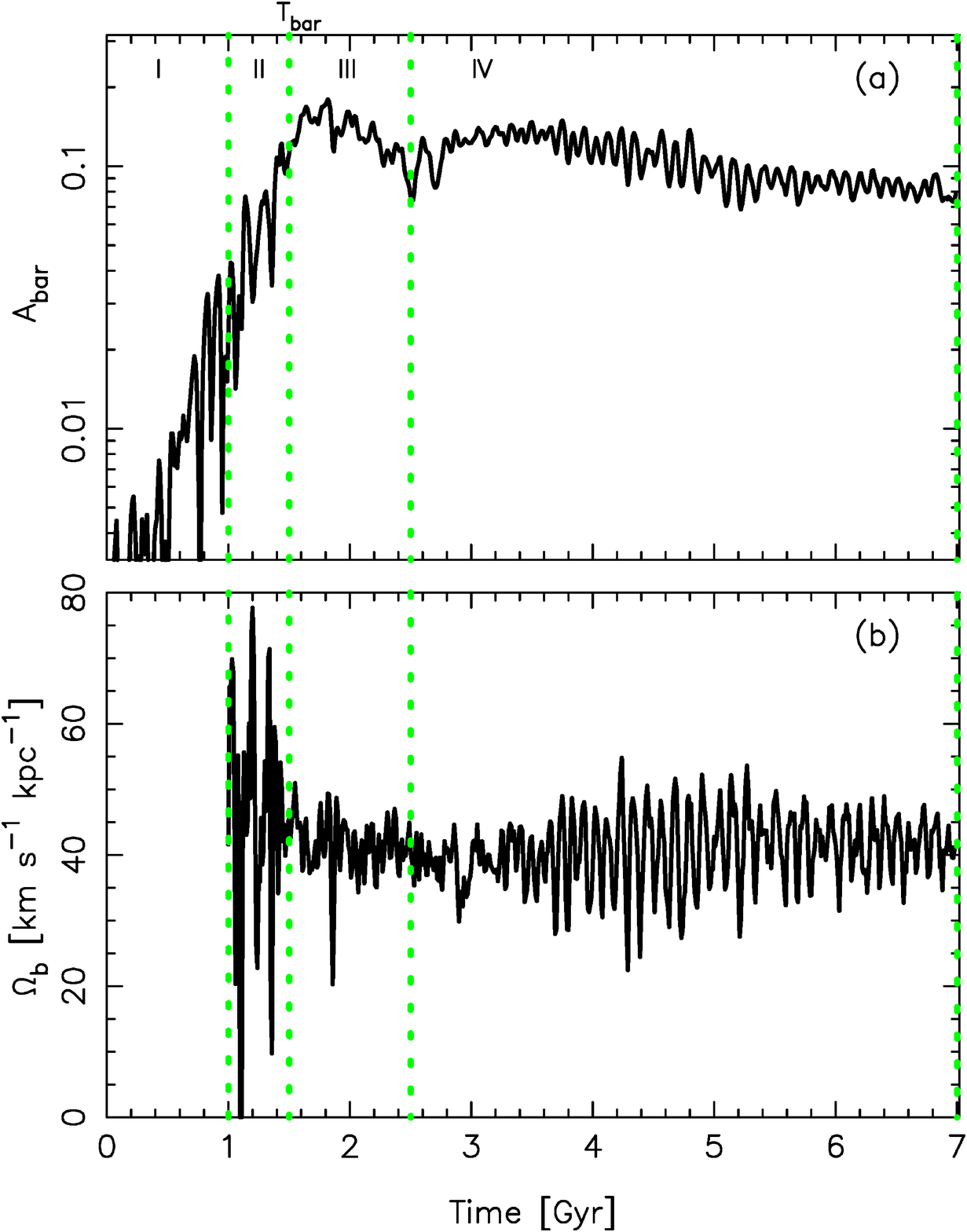}
\caption{
    Time evolution of (a) the bar amplitude, $A_{\rm bar}$, and (b) the bar pattern speed, $\Omega_{\rm b}$, within $R = 3.5$ kpc. 
    The time evolution of $\Omega_{\rm b}$ is not shown for $t < 1$ Gyr, because the bar amplitude is too low to evaluate the pattern speed.
    The green vertical dotted lines indicate the times ($t =$ 1.0, 1.5, 2.5, and 7.0 Gyr) corresponding to those of the snapshots shown in Fig.~\ref{fig:SnapshotEvolution}. 
    These time are chosen to  separate the qualitatively different dynamical phases of the bar, i.e. before bar formation (phase I), bar growing (phase II), just after bar formation (phase III), and bar stabilised (phase IV), as highlighted in panel~(a).
    Throughout this paper, the bar formation time is defined as $T_{\rm bar} = 1.5$ Gyr.
}	
\label{fig:barevol}
\end{center}
\end{figure}

\begin{figure}
\begin{center}
\includegraphics[width=0.45\textwidth]{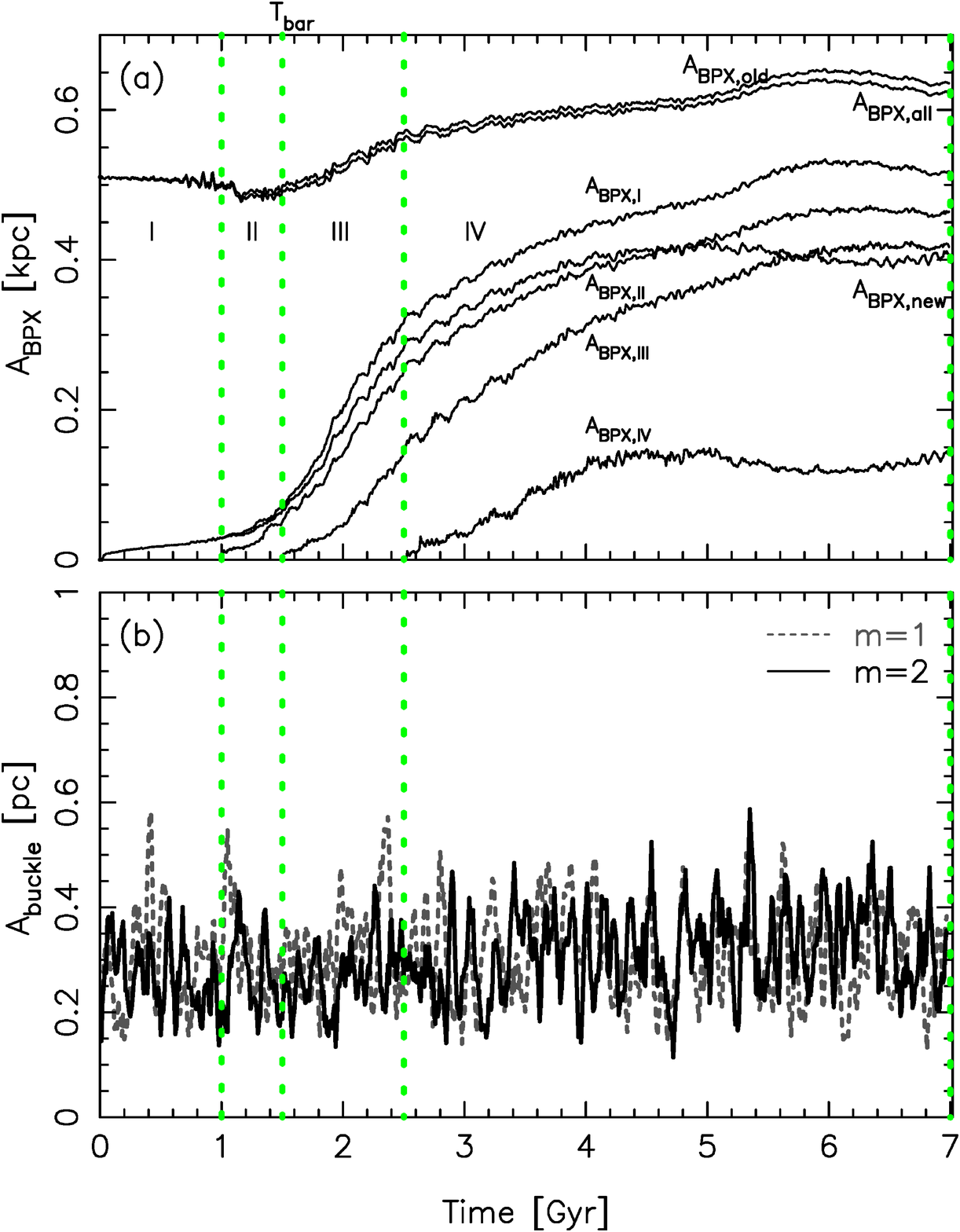}
\caption{
    (a) Time evolution of the BPX {heights}, $A_{\rm BPX}$, measured in $1 < R < 2$ kpc using equation~(\ref{eq:Abpx}). The thicker three solid lines show the time evolution of $A_{\rm BPX}$ for all, old and new stars indicated with the labels of $A_{\rm BPX, all}$, $A_{\rm BPX, old}$ and $A_{\rm BPX, new}$ respectively. The green vertical dotted lines are the same as Fig.~\ref{fig:barevol} and separate the phases I, II, III and IV. The thinner four lines show the time evolution of $A_{\rm BPX}$ for stars born in these four phases, as highlighted with the labels close to the lines.
    (b) 
    Time evolution of the buckling {amplitudes $A_{\rm buckle}$ for $m_z=2$ (solid) and $m_z=1$ (dashed) modes} measured using equation~(\ref{eq:Abuckle}) for the all stars within $1<R<3.5$~kpc. Note that the unit of $y$-axis of panel~(b) is pc.
}	
\label{fig:buckling}
\end{center}
\end{figure}

This section outlines the morphological evolution of the simulated galaxy.
Fig.~\ref{fig:SnapshotEvolution} shows the time evolution (columns) of the simulated galaxy, {with an edge-on view of the stellar distribution along the bar major axis (in the top row), face-on views of the surface densities of stars (in the 2nd row), gas (in the 3rd row) and the star formation rate (in the bottom row).} 
At $t = 1$ Gyr, the bar has not formed yet. Spiral arms are developed, with gaseous filaments distributed along them (middle-left panels). Star-forming regions concentrate along the spiral arms. 
{At $t = 1.5$ Gyr, the bar is fully formed and its size is about 3--4 kpc. Gas streaming towards the nuclear disc is along the leading sides (or so-called offset-ridge) of the bar.}
These feed the intense star-formation in the ring-shaped nuclear gas disc (at $R \lesssim 1$ kpc), as shown in our previous study with the same simulation \citep[][]{BabaKawata2020a}. 
Our bar produces very little star formation, as in its equivalent of the 3 kpc arms of the Milky Way, aside from the bar tips, where also the Milky Way has enhanced star formation \citep[e.g.][]{Veneziani+2017}. While some galaxies do have some star formation along the bar major axis, this desert of star formation is found in many barred galaxies \citep[e.g.][]{MartineFriedli1997,JamesPercival2018}, and strongly suspected in our Milky Way.

To quantify dynamical evolution of the bar, we measure the bar amplitude with the $m=2$ Fourier amplitude of the face-on stellar density maps as
\begin{equation}
 A_{\rm bar} = \left|\frac{\sum_{j=1}^{N} m_j e^{2i\phi_j}}{\sum_{j=1}^{N} m_j}\right|,
 \label{eq:Abar}
\end{equation}
where $m_j$, $\phi_j$ and $N$ are the mass, azimuth angle of a $j$-th stellar particle and the number of stellar particles within a cut-off radius of $R_{\rm c} = 3.5$ kpc, respectively \citep[e.g.][]{SellwoodAthanassoula1986,Dubinski+2009}. Fig.~\ref{fig:barevol}(a) reveals that the bar reaches its maximum amplitude around $t = 1.8$ Gyr, and then its amplitude gradually decreases for 5 Gyr. The exponential growth of the bar strength lasts about {1.5 Gyr}. In the following, we define the bar formation time $T_{\rm bar} = 1.5$ Gyr, when the bar reaches about half its peak amplitude.

The bar pattern speed in our simulation is obtained by calculating time change of the $m=2$ phase ($\phi_{m=2}$) as $\Omega_{\rm b} \equiv \Delta \phi_{m=2}/\Delta t$, where $\Delta t = 10$ Myr.
Fig.~\ref{fig:barevol}(b) shows that the pattern speed is as fast as about 50~\ps{} at $t \simeq 1$ Gyr. Then, it gradually decreases until $t \simeq 1.5$ Gyr, and settles to around 40 \ps{}. The fluctuation of the bar strength and the pattern speed in Fig.~\ref{fig:barevol} {could} indicate that the amplitude and the phase angle of the bar are oscillating \citep{Wu+2018,Hilmi+2020}.

We next investigate the time evolution of the vertical structure of the bar. The top panels of Fig.~\ref{fig:SnapshotEvolution} present the time evolution of the side-on view of the stellar disc. Prior to the bar formation ($t \simeq 1$ Gyr), the bulge region is still elliptical {shaped}. Interestingly, a weak BPX-shaped bulge appears just after bar formation (at $t \simeq 1.5$ Gyr) and continuously sharpens with time. We quantify the BPX {height} by:
\begin{equation}
 A_{\rm BPX} = \left(\frac{\sum_{j=1}^{N} m_j z_j^2}{\sum_{j=1}^{N} m_j}\right)^{1/2},
\label{eq:Abpx}
\end{equation}
where $z_j$ is the vertical position of the $j$-th stellar particle. This is basically a root square mean height of the central bar region. 
We call star particles from the initial condition `old stars', and stars born during the active simulation `new stars'.
Fig.~\ref{fig:buckling}(a) shows that the BPX {heights} for the old ($A_{\rm BPX,old}$), new ($A_{\rm BPX,new}$) and all ($A_{\rm BPX,all}$) stars
start to increase around $t = T_{\rm bar}$ and the growth continues even after the bar fully developed around $t = 2$~Gyr. 
The old stellar component is thicker from the beginning, but thickens further. However, this thickening is less prominent, compared to the new stellar component, which forms the BPX-shaped bulge just after the bar formation.
The strength of the BPX-shape becomes almost constant after $t=4$~Gyr.

As seen in the snapshots of Fig.~\ref{fig:SnapshotEvolution}, the BPX-shaped bulge in our simulation is not developed by bar buckling. To show it more quantitatively, we analyse the temporal evolution of the buckling amplitude, $A_{\rm buckle}$, which is defined by the following equation \citep[e.g.][]{Debattista+2006},
\begin{equation}
 A_{\rm buckle} = \left|\frac{\sum_{j=1}^{N} z_j m_j e^{i m_z\phi_j}}{\sum_{j=1}^{N} m_j}\right|,
\label{eq:Abuckle}
\end{equation}
{for the $m_z$-th mode buckling.}
As shown in Fig.~\ref{fig:buckling}(b), $A_{\rm buckle}$ {for the both $m_z=1$ and $2$ modes} remains less than 1 pc at all the time. This is much smaller than the gravitational softening length, and confirms that our BPX-shaped bulge is not caused by bar buckling. 

{
It is worth noting that previous studies with $N$-body/hydrodynamics simulations showed that including the gas component suppresses bar buckling, but a thickened bulge appeared \citep[e.g.][]{Debattista+2006,Berentzen+2007,Seo+2019}. However, these simulations do not include star formation, and hence lack the new-born and thus vertically cold stellar populations. Our high-resolution self-consistent simulation shows that the BPX-shaped bulge appears without bar buckling. 
}

\section{Age and Birth Radius Distributions of Bar/Bulge Stars}
\label{sec:Rbirth_tbirth}

\begin{figure*}
\begin{center}
\includegraphics[width=0.95\textwidth]{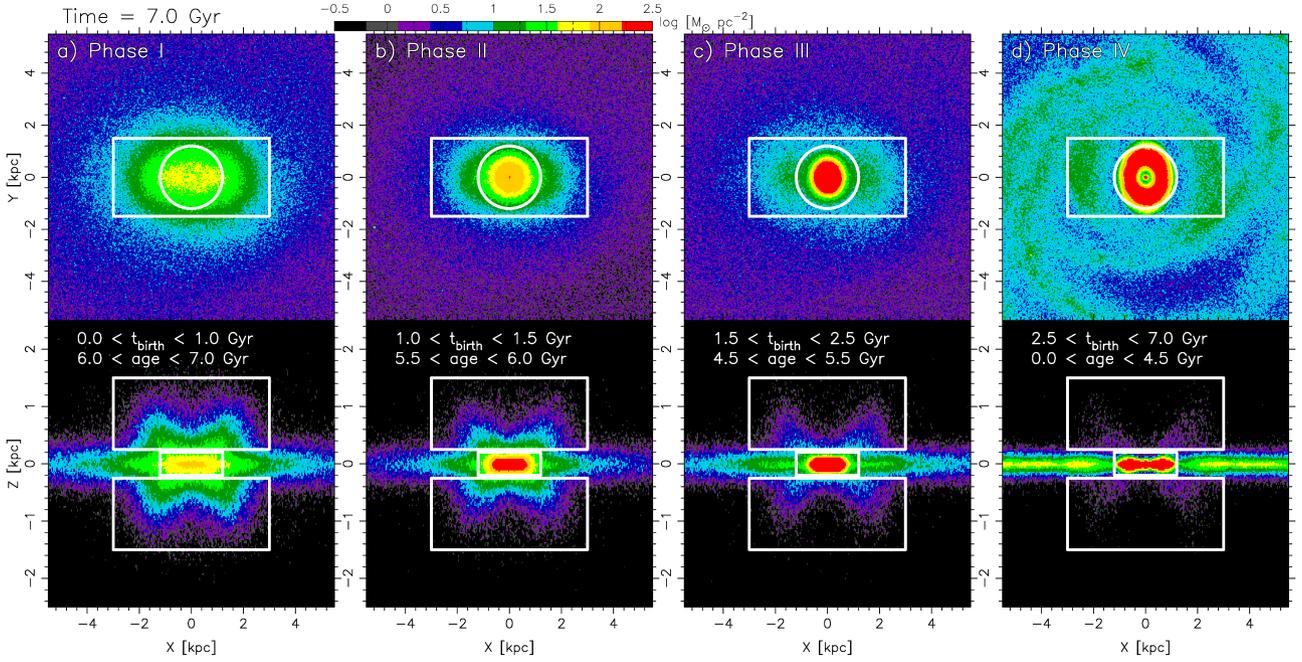}
\caption{
	Face-on (upper) and side-on (lower) spatial distributions of stars with various age ranges at $t = 7$ Gyr. From left to right, the panels show the distribution of stars formed 
	a) before bar formation, 
	b) in the bar growing phase, 
	c) after bar formation, 
	d) in the bar stable phase, corresponding to phases~I, II, III and IV in Fig.~\ref{fig:barevol}, respectively. 
	The white circles and rectangles in each panel highlights the NSD region 
	and the BPX region, respectively.
}	
\label{fig:SnapshotMAP}
\end{center}
\end{figure*}

Fig.~\ref{fig:SnapshotMAP} shows the face-on (upper panels) and edge-on (lower panels) stellar density distributions of four different age populations at $t=7$~Gyr. 
The four populations are delimited by the green vertical dotted lines in Figs.~\ref{fig:barevol} and \ref{fig:buckling}, and indicated with phases I, II, III and IV. 
We can see that the stars formed in phases~I and II, i.e. $0 < t_{\rm birth} < T_{\rm bar}$ Gyr, show strong BPX-shape when viewed edge-on. This is also seen as the high BPX amplitudes of the stars formed in these phases ($A_{\rm BPX,I}$ and $A_{\rm BPX,II}$ in  Fig.~\ref{fig:buckling}(a)).
The BPX-shaped bulge region is less populated by the stars formed after bar formation, $t_{\rm birth}>T_{\rm bar}$~Gyr (i.e. phases~III and IV). Fig.~\ref{fig:buckling}(a) shows that the BPX amplitude of stars formed in phase~III, $A_{\rm BPX,III}$, is still high. However, that of stars formed in phase~IV, $A_{\rm BPX,IV}$, is significantly weaker.

We next analyse the birth time ($t_{\rm birth}$) and birth radius ($R_{\rm birth}$) distributions of the stars in the different volumes in the bar regions. To this end,
we first divide the bar/bulge stars at $t = 7$ Gyr into three groups, namely the NSD, the BPX-shaped bulge (BPX hereafter) and the long bar (BAR hereafter). The NSD region is defined as a cylindrical region of $R < 1.2$ kpc and $|z| < 0.2$ kpc (see Fig.~\ref{fig:SnapshotMAP}). The BPX region is defined as $|x| < 3$ kpc, $|y| < 1.5$ kpc and $|z| > 0.25$ kpc. The BAR region is defined as $|x| < 3$ kpc, $|y| < 1.5$ kpc, $|z| < 0.2$ kpc and $R > 1.5$ kpc, and the BAR stars are restricted with $|v_{\rm z}|< 45$~km~s$^{-1}$, which is comparable to the velocity dispersion in $|z|<0.2$ kpc.
Then, we traced these particles backward in time and determined the birth radius of the star particle, $R_{\rm birth}$ {using the snapshots saved every 1 Myr. Since the radial velocity dispersion of the stars is about $10~\rm km~s^{-1}$ at the birth time, this time resolution is sufficient to estimate the birth radii of the stars with an accuracy of 10 pc.} 
{Note that we neglected the stars which were present at $t=0$ in the following analysis. We found that these stars occupy about 95\% and 70\% of the BPX and BAR populations, respectively. 
Although these stars are dominant, they are regarded as the older stars. Since the focus of this study is on the youngest limits of the age distributions of BPX stars, the fraction of the pre-existing stars prior to the bar formation does not affect the following discussion.
}

\begin{figure*}
\begin{center}
\includegraphics[width=0.95\textwidth]{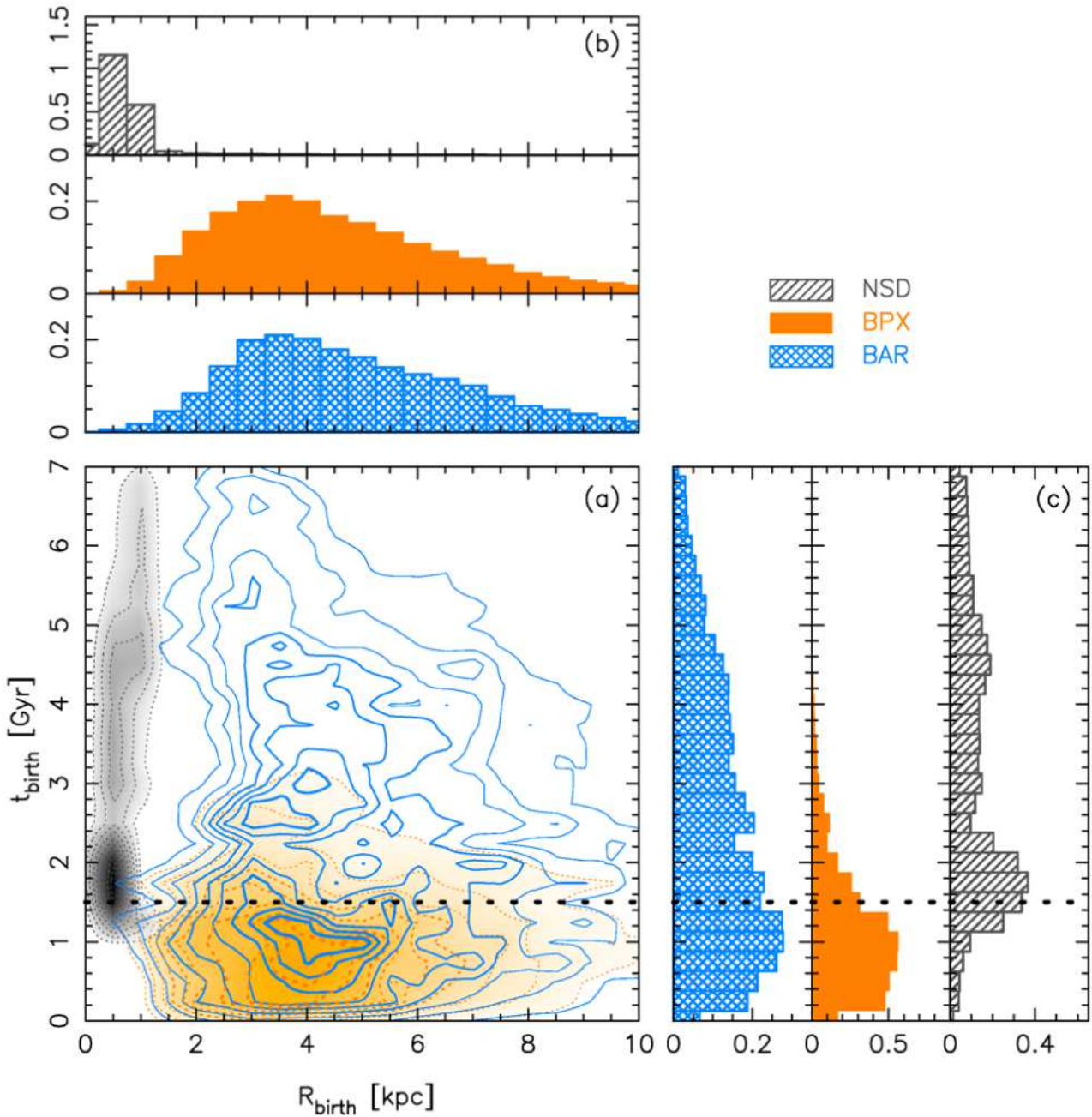}
\caption{
	(a) $R_{\rm birth}$ versus $t_{\rm birth}$ density map of the NSD (gray, dashed), BPX (orange, dashed), BAR (blue, solid) stars at $t = 7$ Gyr. 
	The horizontal dashed line indicates the bar formation time $T_{\rm bar}$ (=1.5 Gyr).
	(b) $R_{\rm birth}$ distributions of the NSD (upper), BPX (middle), and BAR (lower) samples.
	(c) $t_{\rm birth}$ distributions of the NSD (right), BPX (middle), and BAR (left) samples.
}
\label{fig:AgeRbirthDistribution}
\end{center}
\end{figure*}

\begin{figure}
\begin{center}
\includegraphics[width=0.45\textwidth]{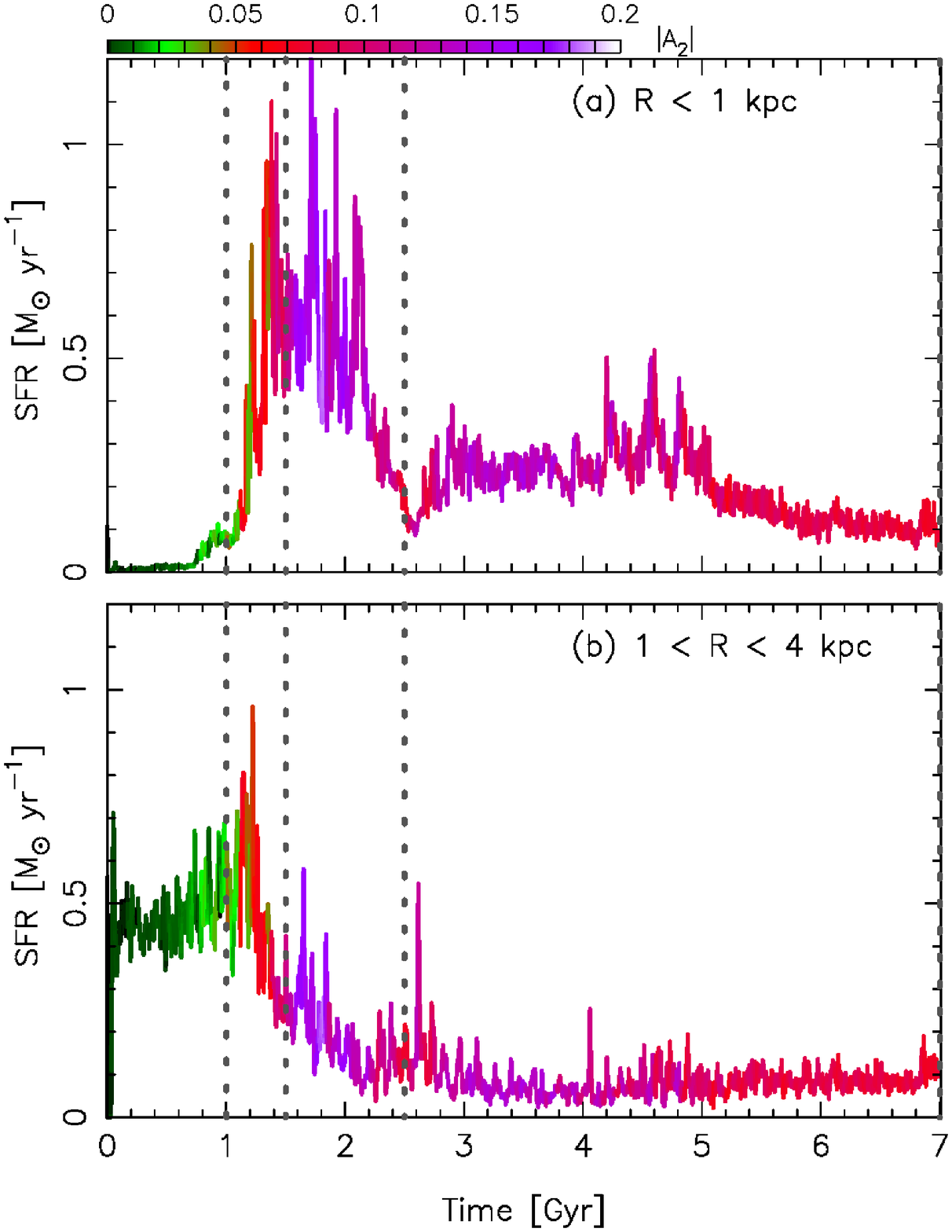}
\caption{
	Time evolution SFR within (a) $R<1$~kpc and (b) $1<R<4$~kpc. The vertical dashed lines indicate the times corresponding to those of the snapshots shown in Fig.~\ref{fig:SnapshotEvolution}. Colors indicate the bar amplitude, $|A_{\rm bar}|$, defined in eq.~(\ref{eq:Abar}).
}
\label{fig:SFR}
\end{center}
\end{figure}

Fig.~\ref{fig:AgeRbirthDistribution}(a) displays that the above three groups are distributed in the $R_{\rm birth}$--$t_{\rm birth}$ plane differently. 
The NSD stars (gray dashed contours) are formed almost exclusively at $t_{\rm birth} \gtrsim 1 $ Gyr.
We plot the temporal evolution of the in-situ SFRs of the central region in Fig.~\ref{fig:SFR}(a). The colour of the line indicates the bar amplitude. The figure shows that when the bar starts forming around $t=1$~Gyr, the high level of star formation in the central region is triggered and continues until $t\sim2$~Gyr, followed by the continuous low-level of star formation.
Consequently, the $t_{\rm birth}$-distribution of the NSD stars in Fig.~\ref{fig:AgeRbirthDistribution}(c) shows a peak around $1.8$ Gyr with a long tail until $t_{\rm birth} = 7$ Gyr.
Fig.~\ref{fig:AgeRbirthDistribution}(b) demonstrates that the $R_{\rm birth}$-distribution of the NSD stars sharply peaks around 0.5 kpc and almost all stars originate from $R_{\rm birth}<1.5$ kpc. Vice versa all stars formed in the NSD region after $t > 1$ Gyr remain confined to this region.

In contrast to the NSD stars, the BPX stars (orange dotted contours) are dominated by the population with $t_{\rm birth} \lesssim 1.5 $ Gyr formed at $1~{\rm kpc}\lesssim R_{\rm birth} \lesssim 4~{\rm kpc}$ (Fig.~\ref{fig:AgeRbirthDistribution}(a)). In other words, the formation time and radius of the BPX stars are separated from those of the NSD stars (Figs.~\ref{fig:AgeRbirthDistribution}(b) and (c)). 

The distribution of the BAR stars (blue solid contours) in Fig.~\ref{fig:AgeRbirthDistribution} is similar to that of the BPX stars. However, 
a significant fraction of the BAR stars formed after $t_{\rm birth}\approx1.5$~Gyr. 
The $R_{\rm birth}$ distribution of the BAR stars with $t_{\rm birth}\gtrsim1.5$~Gyr shows that they formed in the outer region of the bar, including outside the bar ($R>3$~kpc),
and fell into the bar region later, because the star formation in the bar is quenched inside the bar, except the central NSD region (Fig.~\ref{fig:SFR}(b)). 
These differences in $t_{\rm birth}$ and $R_{\rm birth}$ between the BPX and BAR stars suggest that the efficiency for stars to be in the BPX-shaped bulge is higher for stars formed in the inner disc, $R \lesssim 4$~kpc, before the bar fully formed.

\section{Orbital properties of Bar and BPX-shaped bulge stars}
\label{sec:Orbit}

\begin{figure}
\begin{center}
\includegraphics[width=0.45\textwidth]{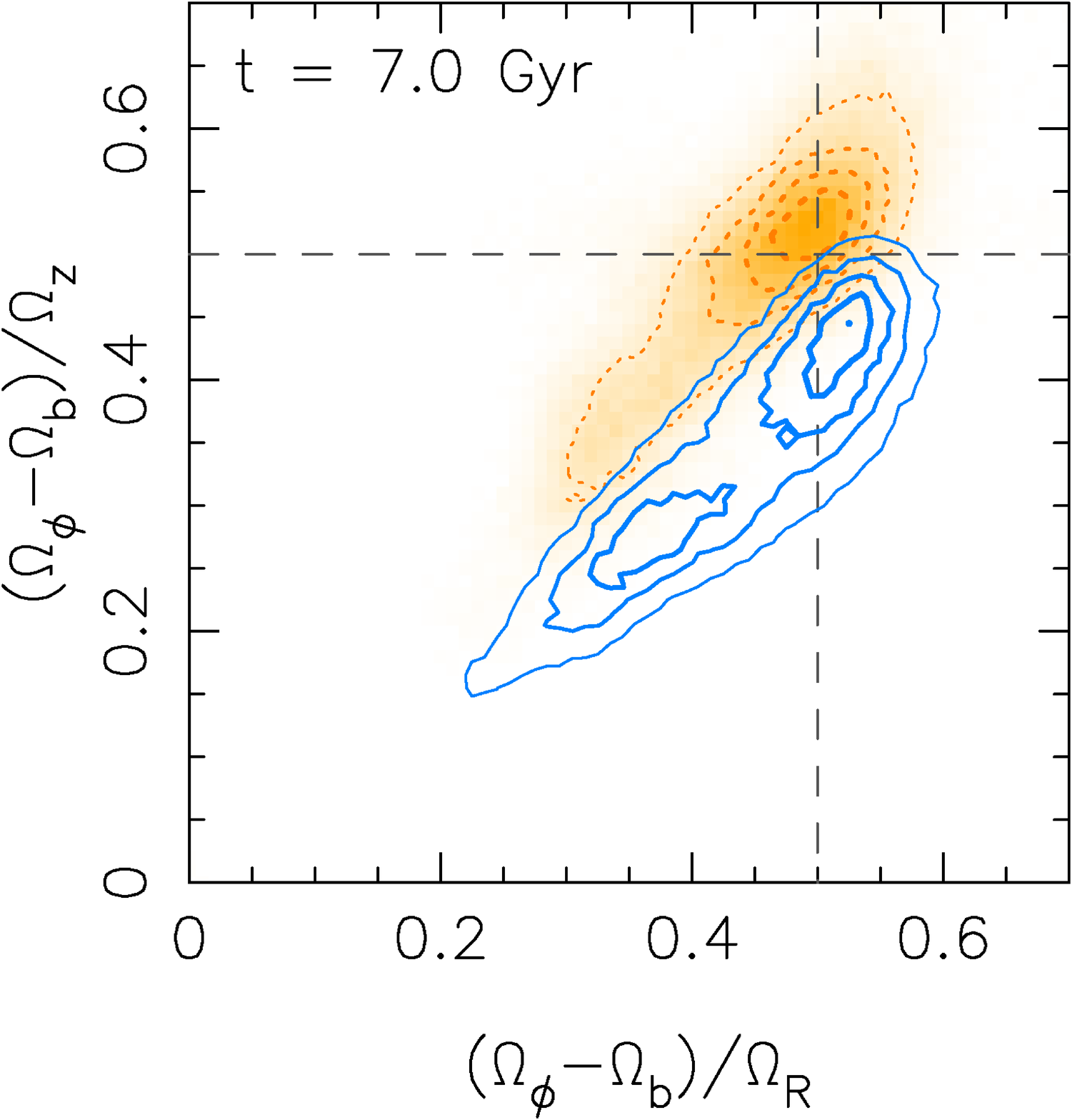}
\caption{
Density maps of orbital frequencies of the BPX (orange dashed) and BAR (blue solid) stars with $0<t_{\rm birth}<1.5$~Gyr at $t=7.0$ Gyr.
Contour levels are 0.1, 0.3, 0.5, 0.7, and 0.9 of the peak.
}	
\label{fig:OrbitFreq}
\end{center}
\end{figure}

\begin{figure*}
\begin{center}
\includegraphics[width=0.95\textwidth]{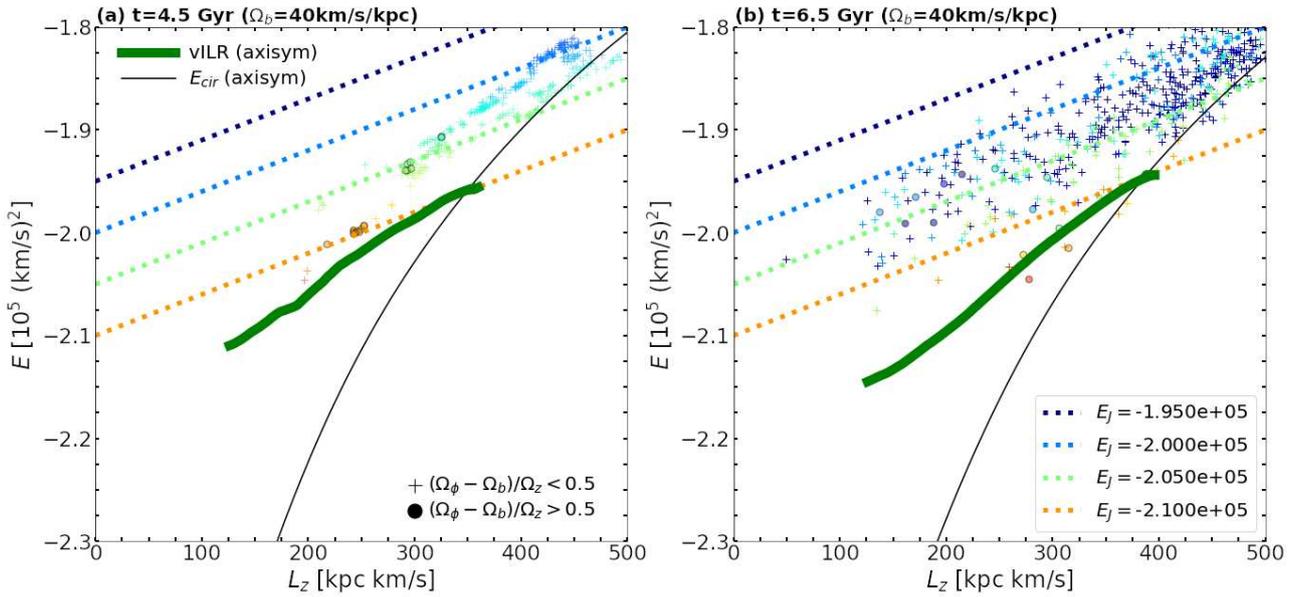}
\caption{
The distribution of $L_{z}$ and $E$ at (a) $t=4.5$~Gyr and (b) $t=6.5$~Gyr for stars ({crosses and circles}
) formed at $4.47 < t_{\rm birth} < 4.50$ Gyr, but not within the NSD ($R<1.5$~kpc and $z<0.2$~kpc). 
The thin solid curve indicates $E_{\rm c}(L_z)$ which is orbital energy expected for circular orbits in the axisymmetric potential approximated from the simulation snapshot.
{
Because the stars' $E$ are computed from the full non-axisymmetric potential, the stars can be distributed slightly below $E_{\rm c}(L_{\rm z})$.
}
Green thick line indicate the location of the axisymmetric vILR, which is estimated in the axisymmetric potential using {\tt AGAMA}.
{The stars with $(\Omega_\phi-\Omega_{\rm b})/\Omega_{\rm z}<0.5$ are marked with crosses and the stars with $(\Omega_\phi-\Omega_{\rm b})/\Omega_{\rm z}>0.5$ are marked with circles. The colours of the symbols are given according to the value of $E_{\rm J}$ at $t=4.5$ Gyr.
}
Dotted lines show constant Jacobi energy ($E_{\rm J}$ in $\rm km^2~s^
{-2}$) {at the time of each panel} every $0.05 \times 10^5~\rm km^2~s^{-2}$. 
{
There are more stars in panel (b), because more stars outside of panel (a) fall into the plot range at $t=6.5$~Gyr.
}
}
\label{fig:Ej4.5Gyr}
\end{center}
\end{figure*}

\begin{figure*}
\begin{center}
\includegraphics[width=0.95\textwidth]{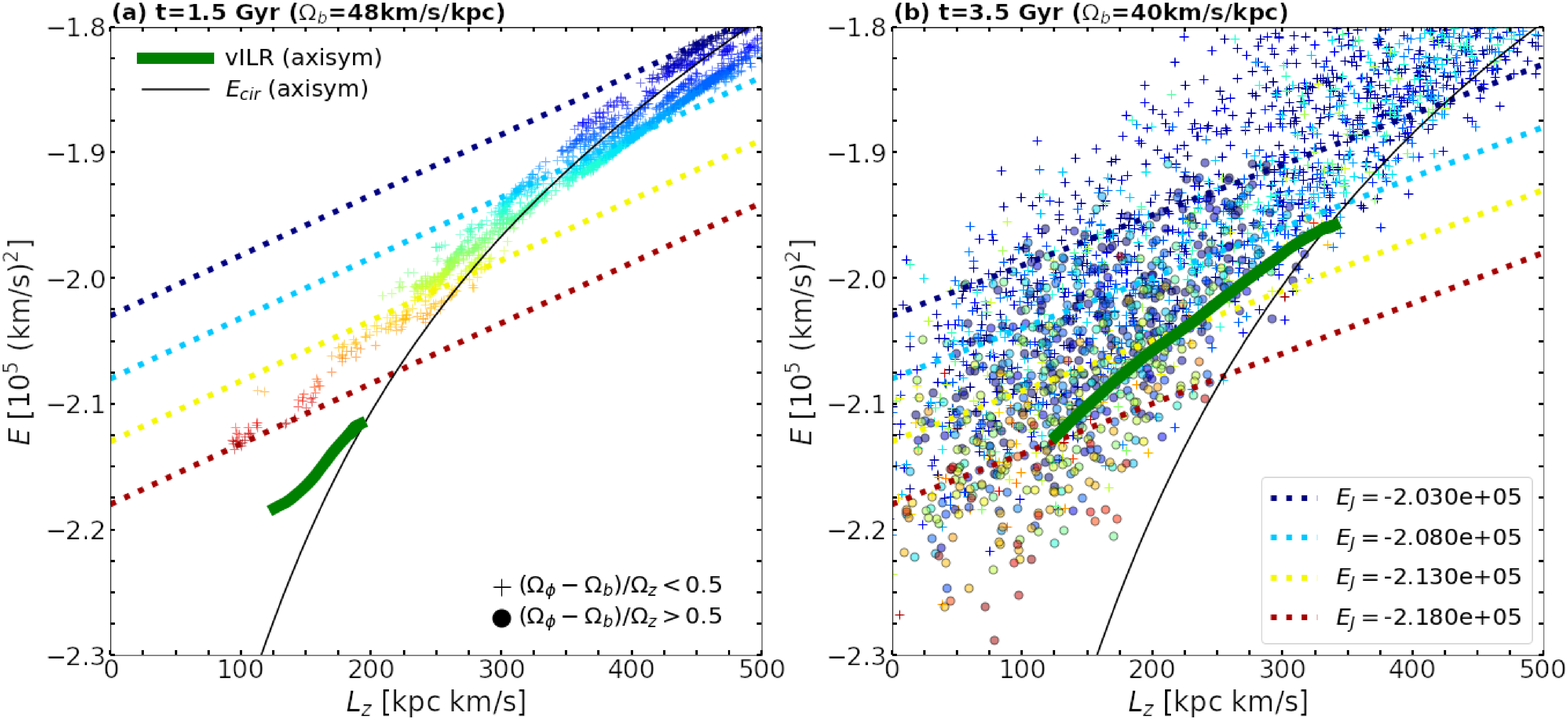}
\caption{
Same as Fig.~\ref{fig:Ej4.5Gyr}, but for stars formed at $1.47 < t_{\rm birth} < 1.50$ Gyr. 
$\Omega_{\rm b}=48$ and $40~\rm km~s^{-1}~kpc^{-1}$ is assumed {at $t=1.5$ and 3.5~Gyr, respectively}.
}
\label{fig:Ej1.5Gyr}
\end{center}
\end{figure*}

To study the mechanism causing the differences in the $t_{\rm birth}$ and $R_{\rm birth}$ distributions between the BAR and BPX stars as described above, we analyse orbital properties in these two components. 
For this analysis, we follow the method of \citet{SellwoodGerhard2020} and evaluate the orbital frequencies (radial frequency $\Omega_R$, azimuthal frequency $\Omega_{\phi}-\Omega_{\rm b}$, and vertical frequency $\Omega_z$) 
{
at time $t$ from the previous 300 Myr of evolution, with outputs every 1 Myr.
}

We first investigate the mechanism to form the BPX-shaped bulge without bar buckling. To this end, we applied the frequency analysis to the stars that were born at $0<t_{\rm birth}<1.5$~Gyr and ended up either in the BPX or BAR region at $t=7$~Gyr. 
Fig.~\ref{fig:OrbitFreq} shows the distributions of the BPX (orange dashed contours)
and BAR (blue solid contours) stars in the $(\Omega_{\rm \phi}-\Omega_{\rm b})/\Omega_{\rm R}$ vs.\ $(\Omega_{\rm \phi}-\Omega_{\rm b})/\Omega_{\rm z}$ measured.
The horizontal dashed line indicates the vILR, $(\Omega_{\rm \phi}-\Omega_{\rm b})/\Omega_{\rm z}=0.5$, and the vertical dashed line shows the hILR, $(\Omega_{\rm \phi}-\Omega_{\rm b})/\Omega_{\rm R}=0.5$. It is interesting to see that the majority of the BPX stars have \ $(\Omega_{\rm \phi}-\Omega_{\rm b})/\Omega_{\rm z}>0.5$. On the other hand, there are almost no BAR stars which reach $(\Omega_{\rm \phi}-\Omega_{\rm b})/\Omega_{\rm z}>0.5$. This contrast infers that the BPX-shaped bulge is built up by vILR heating, as discussed in previous studies \citep[e.g.][]{Combes+1990,Quillen+2014,WozniakMichel-Dansac2009,SellwoodGerhard2020}.
Hence, the stars need to reach the vILR to be heated up to the BPX-shaped bulge. In general, the vILR is in the inner disc, and therefore it is easier for the stars formed in the inner disc before the bar formation to be heated to the BPX region. As discussed above, after the bar formation ($t \gtrsim 1.5$ Gyr), the star formation is suppressed in the bar region, except in the NSD (Fig.~\ref{fig:SFR}). 
Although the stars continue forming in the outer region of the bar, including outside the bar, and fall into the inner region of the bar, Fig.~\ref{fig:AgeRbirthDistribution} shows that such stars remain in the BAR region.

To investigate why the stars formed in the outer disc after the bar formation are less affected by the vertical heating, {we analyse the distribution of angular momentum, $L_{\rm z}$, and total orbital energy {(i.e. particle's total kinetic energy plus gravitational potential)}, $E$,
at $t=4.5$ and 6.5~Gyr for stars with $4.47 < t_{\rm birth} < 4.50$~Gyr}, i.e. formed well after the bar formation, but formed outside the NSD  (Fig.~\ref{fig:Ej4.5Gyr}). {We use $L_{\rm z}$ and $E$ {rather} than actions, because actions are difficult to compute under the strong non-axisymmetric potential, like in the bar region, but $L_{\rm z}$ and $E$ can be computed from the position and velocity of stars and the gravitational potential at the location of the stars. We compute the gravitational potential from the full particle distribution of a snapshot output of the simulation, i.e. taking into account the bar potential shape.} We also compute Jacobi energy, $E_{\rm J}~(\equiv E-\Omega_{\rm b}L_{\rm z})$, with the bar pattern speed of $\Omega_{\rm b} = 40~{\rm km~s^{-1}~kpc^{-1}}$ at both $t=4.5$ and 6.5~Gyr, which are shown with the {dotted lines} in this figure. 

Fig.~\ref{fig:Ej4.5Gyr}(a) shows that since most of these stars formed in the outer region of the bar, they have {higher $L_{\rm z}$ and} $E_{\rm J}$ at $t=4.5$~Gyr, i.e. just after they formed. Fig.~\ref{fig:Ej4.5Gyr}(b) shows that some of these stars lose the angular momentum after 2~Gyr.
Because they conserve $E_{\rm J}$ and move along the {dotted} lines of constant $E_{\rm J}$, which is expected under the bar potential with steady amplitude and constant pattern speed \citep[e.g.][]{BinneyTremaine2008,Chiba+2021}, their $E$ also changes and deviates more from $E_{\rm c}(L_{\rm z})$, which is the {total orbital} energy for circular orbits in the average axisymmetric potential as a function of $L_{\rm z}$. {Conservation of $E_{\rm J}$ is shown in Fig.~\ref{fig:Ej4.5Gyr}(b), where the symbols are coloured with their $E_{\rm J}$ at $t=4.5$~Gyr, and the dotted lines indicate $E_{\rm J}$ at $t=6.5$~Gyr. Similarity in colour between the symbols and the dotted lines close to them indicate their $E_{\rm J}$ at $t=6.5$~Gyr are similar to that at $t=4.5$~Gyr. We confirm that the majority of stars conserve their $E_{\rm J}$ within 5\%.
 }

In Fig.~\ref{fig:Ej4.5Gyr}, we also show the thick green lines of `axisymmetric' vILR. This is a location of vILR computed from an axisymmetric potential approximated from the real potential.
{Strictly speaking, this is not correct vILR in the non-axisymmetric system. However, we use this for an indicator of the location of vILR. 
}
In this study, we use the {\tt AGAMA} software package \citep{Vasiliev2019}
to obtain an approximate `axisymmetric' potential of the each snapshot.
We use the {\tt Multipole} function for the DM halo and classical bulge and the {\tt CylSpline} function for the disc in {\tt AGAMA}. We then computed the orbital frequencies ($\Omega_{\rm R,axi}$, $\Omega_{\rm \phi,axi}$, $\Omega_{\rm z,axi}$) for stars, and the locus of the axisymmetric vILR is drawn from the stars within $0.45 < (\Omega_{\rm \phi,axi}-\Omega_{\rm b})/\Omega_{\rm z,axi} < 0.55$.
We can see that most of the stars in Fig.~\ref{fig:Ej4.5Gyr} cannot reach the vILR, because they keep the high $E_{\rm J}$. 
{As a result, very few stars end up at the BPX region, i.e. $(\Omega_{\rm \phi}-\Omega_{\rm b})/\Omega_{\rm z}>0.5$, which are highlighted with circles.}

$E_{\rm J}$ is only a conserved quantity, when the bar potential does not change with time. Accordingly, we observe that significant amount of stars change $E_{\rm J}$ before the bar potential settles.
{Fig.~\ref{fig:Ej1.5Gyr} shows $L_z$, $E$
and $E_{\rm J}$ at $t=$ 1.5~Gyr and 3.5~Gyr for stars formed at $1.47 < t_{\rm birth} < 1.50$~Gyr.}
{The contours of $E_{\rm J}$ (dotted lines) and the position of vILR are different between Fig.~\ref{fig:Ej1.5Gyr}(a) and (b), because} the gravitational potential {and the pattern speed of the bar change} significantly between 1.5 and 3.5~Gyr. {Differences between colours of symbols and colours of their closest dotted lines in Fig.~\ref{fig:Ej1.5Gyr}(b)
shows that significant number of stars lose $E_{\rm J}$ (up to about 15\%) and reach vILR at $t=3.5$~Gyr.
Consequently, more stars are heated up to the BPX region, i.e. having $(\Omega_{\rm \phi}-\Omega_{\rm b})/\Omega_{\rm z}>0.5$ (circles in Fig.~\ref{fig:Ej1.5Gyr}(b)) at $t=3.5$~Gyr.} 
Hence, until the bar becomes stable, the stars formed outside vILR can lose $E_{\rm J}$, and they can be heated up to the BPX region by the vertical heating at vILR. This explains the tail of the BPX stars with $t_{\rm birth}>1.5$~Gyr in Fig.~\ref{fig:AgeRbirthDistribution}(c). This violation of the conservation of $E_{\rm J}$ during the bar formation also means that stars formed at various radius and orbits can reach vILR and can be heated up to the BPX region, which helps the continuous growth of the BPX-shaped bulge until the bar becomes stable (Fig.~\ref{fig:buckling}).

{In addition to the non-conservation of $E_J$, we find that an outward shift of the vILR location also helps the formation of the BPX-shaped bulge in our simulation. Comparing Fig.~\ref{fig:Ej1.5Gyr}(a) and Fig.~\ref{fig:Ej1.5Gyr}(b), the vILR (in an axisymmetric potential approximation, but as a rough indicator of the position of vILR) is located at $L_{\rm z}\lesssim 200~\rm km~s^{-1}~kpc^{-1}$ at $t=1.5$ Gyr, but it shifted to $L_{\rm z}\lesssim 350~\rm km~s^{-1}~kpc^{-1}$ at $t=3.5$ Gyr. This also helps for more stars to reach the vILR at $t=3.5$~Gyr.
This outward shift of the vILR location is due to an increase in the central mass concentration. 
The bar formation triggers gas inflow to the centre, which increases the central mass concentration and shifts the vILR location outwards, causing the stars to be heated up to the BPX region.}

\section{Summary and Discussion}
\label{sec:Conclusions}

In this study, we analysed a 3D $N$-body/hydrodynamics simulation of a Milky Way-like barred spiral galaxy and investigated the formation of the BPX-shaped bulge without bar buckling. 
We initialised our simulations with parameters that ensure the absence of violent buckling. Yet, in our simulation the BPX-shaped bulge starts appearing just after the bar formation via the vILR heating. 
This contrasts to BPX-shaped bulges created by violent buckling, which typically happens 1--2 Gyrs after bar formation \citep[e.g.][]{Debattista+2020}.

We find a strong dichotomy in age and formation radius between the stars in the NSD and the BPX-shaped bulge. The BPX-shaped bulge is dominated by the stars formed in the inner disc before the bar formation, and is reached by very few stars younger than the bar formation epoch. Thus, the age distribution of the BPX starkly contrast with that of the NSD, which is formed from gas driven inward after the bar formation \citep[e.g.][]{FriedliBenz1995,Athanassoula2005,BabaKawata2020a}. 
On the other hand, the age and formation radius of the BAR stars, i.e. a long bar stars, do not show a clear transition around the bar formation.

This strong dichotomy in the age distribution between the BPX and the NSD is driven by the quenching of star formation within the main bar region and barriers preventing the migration of a sizeable number of stars into the region of the vIRL: while there is continuous star formation in the outer disc, these stars inevitably have higher Jacobi energy, which is approximately conserved. While they can still lose angular momentum to fall towards the inner bar region, constant $E_J$ implies that this needs to be traded against increased {non-circular energy.}
This in turn prevents them from reaching the vILR and so from being heated to the BPX region. We note that Jacobi energy is not conserved during bar formation (due to the time-dependence of the bar potential). This helps stars formed prior to the time at which the bar potential stabilises at various radii to reach vILR and continuously form the BPX-shaped bulge.

To our knowledge, it is novel to {study the age distribution of the BPX-shaped bulge, compared to the formation time of the bar, focusing on the case when the BPX-shaped bulge is built up by the vILR heating without bar buckling, and we find that the BPX-shaped bulge mainly consists of the stars older than the bar formation epoch.} Then, as shown in Fig.~\ref{fig:AgeRbirthDistribution}(c), the age distribution of the NSD and BPX-shaped bulge are exclusive to each other. Hence, if we could observe the relative age difference between the NSD and the BPX-shaped bulge, and found {an epoch when the decreasing older stars in the NSD coincide with decreasing younger stars in} the BPX-shaped bulge, it {implies} that the BPX-shaped bulge of the galaxy is likely to be built up without bar buckling, and the transition age corresponds to the formation epoch of the bar. 

{Note that our simulation includes a classical bulge with a mass of about 15\% of the disc mass, which is somewhat larger than the currently suggested upper limit of the classical bulge mass fraction in the Milky Way \citep[$\lesssim 10$\%; e.g.][]{Shen+2010,Debattista+2017}.
\citet{Li+2015} demonstrated that without an initial central mass concentration (in their case a compact classical bulge), no $x_2$ orbits develop, and the gas inflow due to the bar formation does not trigger the formation of a nuclear gas ring or NSD \citep[see also][]{Athanassoula1992b,Kim+2012b}. If the Milky Way has no classical bulge, then NSD formation may be considerably delayed relative to the time of bar formation, and the age distributions of stars between the BPX-shaped bulge and the NSD may have a gap even if the BPX-shaped bulge formed without bar buckling. 
However, we note that \citet{Li+2015} assumed external bar potentials and ignore self-gravity of the gas. 
Self-consistent $N$-body/SPH simulations of disc galaxies without a classical bulge showed that nuclear star formation occurs immediately after bar formation in a nuclear gas ring region with a radius of about 10 pc, and the size of the ring depends on the mass of the central mass concentration \citep{Seo+2019}. Hence, we think that further studies are required to investigate, how much delay since the bar formation is expected for the NSD to build up, depending on the size of the compact bulge. If there is a delay of the formation of the NSD, the gap of the age distributions between the BPX-shaped bulge and the NSD may tell us a lack of the central mass concentration before the bar formation. We plan to explore such simulations in a future study. 
}

{
It is interesting to compare our results with the simulations of the BPX-shaped bulge formed via bar buckling. Our study demonstrates that the $N$-body/hydrodynamics simulations with self-consistent star formation model are important to study the age distribution of stars in the different components of the galactic bar, compared with the bar formation epoch, because it depends on the suppression and/or enhancement of star formation due to the bar formation at the different location of the galactic disc. As mentioned in Section~\ref{sec:Introduction}, the presence of gas components suppresses buckling of bars, as shown in previous $N$-body/hydrodynamics simulations \citep[e.g.][]{Berentzen+2007,IannuzziAthanassoula2015}. Hence, it may be difficult to set up a controlled simulation which leads to bar buckling, but including the gas component, radiative cooling and self-consistent star formation model. Interestingly, \citet{Fragkoudi+2020} reports that some of the Milky Way-like galaxies in the Auriga high-resolution cosmological simulations underwent the bar buckling. It would be interesting to compare the age distribution of the stars in the BPX-shaped bulge in their simulations, and study how it depends on the strength of bar buckling. 
}

The age distribution difference between the NSD and the BPX-shaped bulge in the extra-galaxies can be observed with the Integral Field Units (IFUs), like the TIMER survey \citep{Gadotti+2015,Gadotti+2019}. In the Milky Way, the age and orbit of giant stars in the BPX-shaped bulge can be observed with the near-infrared (NIR) multi-objects spectrograph, like APOGEE \citep[e.g.][]{Bovy+2019,Queiroz+2021,Wylie+2021}.
However, it is challenging to observe the age of the stars in the NSD. This may require to use bright stars, like Mira variables \citep{Matsunaga+2009}, which are known to follow the age-period relation \citep{Feast+2006,Grady+2019,Grady+2020}. {\it Japan Astrometry Satellite Mission for INfrared Exploration} \citep[{{\it JASMINE};}][]{Gouda2012,Gouda+2020}\footnote{
\url{http://jasmine.nao.ac.jp/index-en.html}} will provide the NIR astrometry of the stars in the Galactic central region, including Miras in the NSD, and provide the accurate measurement of the transverse velocity of the NSD stars. The combination of these future observations will enable us to test if the Milky Way's BPX-shaped bulge is formed by the vILR heating without buckling, and reveal the bar formation epoch.

\section*{Acknowledgements}

{We thank the anonymous referee for his/her constructive and helpful comments which have improved the manuscript. }
We thank Takayuki R. Saitoh for technical supports on performing numerical simulations with {\tt ASURA}. Calculations, numerical analyses and visualization were carried out on Cray XC50 (ATERUI-II) and computers at Center for Computational Astrophysics, National Astronomical Observatory of Japan (CfCA/NAOJ). This work was supported by the Japan Society for the Promotion of Science (JSPS) Grant-in-Aid for Scientific Research (C) Grant Numbers 18K03711 and 21K03633. JB acknowledges the supports by JSPS KAKENHI grant Nos. 17H02870, 18H01248, 19H01933 and 21H00054. DK acknowledges the support of the UK's Science \& Technology Facilities Council (STFC Grant ST/N000811/1). RS acknowledges the generous support of a Royal Society University Research Fellowship.

\section*{Data Availability}

The data underlying this article will be shared on reasonable request to the corresponding author.

\bibliographystyle{mn2e}

\end{document}